\documentclass[aip,amsmath,amssymb,reprint]{revtex4-2}
\usepackage{xcolor}
\usepackage{graphicx}
\usepackage{hyperref}
\usepackage{dcolumn}
\usepackage{bm}
\usepackage{etoolbox} 
\usepackage{braket}
\usepackage[english]{babel}
\makeatletter
\def\@email#1#2{%
 \endgroup
 \patchcmd{\titleblock@produce}
  {\frontmatter@RRAPformat}
  {\frontmatter@RRAPformat{\produce@RRAP{*#1\href{mailto:#2}{#2}}}\frontmatter@RRAPformat}
  {}{}
}%
\makeatother

\usepackage{hyperref}

\begin{document}

\title{Parity violation effects in helical osmocene: theoretical analysis and experimental prospects}
\author{Eduardus}
\affiliation{Van Swinderen Institute for Particle Physics and Gravity (VSI), University of Groningen, Groningen, The Netherlands.}

\author{Agathe Bonifacio}
\affiliation{Laboratoire de Physique des Lasers, CNRS, Universit\'{e} Sorbonne Paris Nord, Villetaneuse, France.}

\author{Mathieu Manceau}
\affiliation{Laboratoire de Physique des Lasers, CNRS, Universit\'{e} Sorbonne Paris Nord, Villetaneuse, France.}

\author{Naoya Kuroda}
\affiliation{Department of Micro Engineering, Kyoto University, Kyoto 615-8540, Japan.}

\author{Masato Senami}
\affiliation{Department of Micro Engineering, Kyoto University, Kyoto 615-8540, Japan.}

\author{Juan J. Aucar}
\affiliation{Instituto de Modelado e Innovaci\'on Tecnol\'ogica (UNNE-CONICET), Facultad de Ciencias Exactas y Naturales y Agrimensura,
Universidad Nacional del Nordeste, Av. Libertad 5460, Corrientes, Argentina.}

\author{I. Agust\'{\i}n Aucar}
\affiliation{Van Swinderen Institute for Particle Physics and Gravity (VSI), University of Groningen, Groningen, The Netherlands.}
\affiliation{Instituto de Modelado e Innovaci\'on Tecnol\'ogica (UNNE-CONICET), Facultad de Ciencias Exactas y Naturales y Agrimensura,
Universidad Nacional del Nordeste, Av. Libertad 5460, Corrientes, Argentina.}

\author{Marit R. Fiechter}
\affiliation{Department of Chemistry and Applied Biosciences, ETH Z\"urich, 8093 Z\"urich, Switzerland.}

\author{Trond Saue}
\affiliation{Laboratoire de Chimie et Physique Quantiques, UMR 5626 CNRS -- Universit\`e de Toulouse, 31062 Toulouse Cedex 09, France.}

\author{Jeanne Crassous}
\affiliation{Universit\'{e} de Rennes, CNRS, ISCR-UMR 6226, Campus de Beaulieu, 35042 Rennes Cedex, France.}

\author{Beno\^{i}t Darqui\'{e}}
\affiliation{Laboratoire de Physique des Lasers, CNRS, Universit\'{e} Sorbonne Paris Nord, Villetaneuse, France.}

\author{Shirin Faraji}
\affiliation{Zernike Institute for Advanced Materials, University of Groningen, Groningen, The Netherlands}
\affiliation{Institute of Theoretical and Computational Chemistry, Heinrich Heine University D\"usseldorf, Universit\"atsstra\ss{}e 1, 40225 D\"usseldorf, Germany.}

\author{Luk\'a\v{s} F. Pa\v{s}teka}
\affiliation{Van Swinderen Institute for Particle Physics and Gravity (VSI), University of Groningen, Groningen, The Netherlands.}
\affiliation{Department of Physical and Theoretical Chemistry, Faculty of Natural Sciences, Comenius University, Bratislava, Slovakia.}

\author{Anastasia Borschevsky}
\thanks{\textbf{Author to whom correspondence should be addressed:} a.borschevsky@rug.nl}
\affiliation{Van Swinderen Institute for Particle Physics and Gravity (VSI), University of Groningen, Groningen, The Netherlands.}

\date{\today}

\begin{abstract}
We present a computational investigation of the parity-violating (PV) contributions to the vibrational transitions and nuclear magnetic resonance shieldings of helical osmocene. A number of promising transitions within the spectral window of currently available sub-Hz metrology-grade lasers are identified, exhibiting high intensities and parity violation shifts of up to 7 Hz. We discuss the prospects for the synthesis of this compound and for subsequent ultra-precise mid-IR spectroscopy towards the first detection of parity violation in a chiral molecule. 
\end{abstract}

\maketitle

\section{\label{sec:level1}Introduction}

Chirality has a crucial effect on biological processes in living organisms. One of the greatest unresolved questions in biology is the emergence of biohomochirality: in almost all known biological systems, the most important building blocks, amino acids and sugars, occur naturally only in one enantiomer (D-sugars and L-amino acids), while the other enantiomer is biologically incompatible.\cite{Mason1988,Bonner1995,Blackmond2020}
One of the hypotheses that aims to resolve the mystery of biohomochirality \cite{Cintas2022,Crassous2022} proposes that this is due to a tiny parity-violating energy difference between the two enantiomers. 
This energy difference may have resulted in a tiny excess of one enantiomer over the other in an originally racemic (equal-handed) prebiotic soup, which, amplified over time, may have led to dominance of one of the enantiomers over the other.\cite{YAMAGATA1966495,Mason,Tranter} While this hypothesis is heavily debated,\cite{quack1,wesendrup} the existence of molecular parity violation itself has a firm foundation in particle physics, which predicts that the weak interactions between the electrons and the quarks inside the nuclei should result in a small energy difference between the two enantiomers.\cite{Rein1974,Letokhov75} Parity violation was successfully demonstrated  in nuclear \cite{wu1957} and atomic  experiments.\cite{Bouchiat:2011fs,safronova} Unfortunately, this is not the case with molecular parity violation, despite many years of attempts using a variety of experimental techniques.\cite{daussy,chardonnet:newexp,CraChaSau05,DarStoZri10,Quack2022}

The only experiments so far to have set a strict upper bound on PV effects in chiral molecules were performed at the Laboratoire de Physique des Lasers (LPL) in Paris,\cite{daussy,mason2} using ultra-precise mid-infrared spectroscopy experiments on rovibrational transitions. These measurements were performed on the C--F stretch vibration in the archetypal chiral molecule CHFClBr, setting a limit of $\Delta \nu^{\text{PV}}/ \nu = 2.5\times 10^{-13}$ on the relative PV frequency shift (with $\nu$ the vibrational transition frequency and $\Delta \nu^{\text{PV}}$ the upper limit of the PV frequency difference between the enantiomers, corresponding to about 8 Hz in this case). Subsequent computational studies, however, showed that the actual expected relative PV difference for this transition is a few orders of magnitude smaller ($10^{-17}$) and could not have been reached within the precision of these measurements.\cite{quack3,peter1,thierfelder,rauhut,bruck}

Currently, there is an ongoing attempt to measure PV shifts in vibrational spectra of chiral molecules using an ultrahigh resolution gas-phase spectroscopy experiment based on metrology-grade lasers in the 500--2000$~\text{cm}^{-1}$ range at LPL. \cite{saleh_chiral_2013,saleh_oxorhenium_2018,Cournol_2019} The current set-up is expected to reach relative PV frequency shift sensitivity on the order of $10^{-15}$; therefore, a candidate molecule is needed where the expected PV frequency shift is at least in the order of magnitude of hundreds of mHz to Hz. Additionally, there are practical requirements such a molecule should satisfy -- it should be possible to bring it into the gas phase while maintaining chemical stability, it should ideally be possible to separate it into pure enantiomers, and it should have strong vibrational transitions in a range accessible to current laser technologies. Furthermore, such a molecule should contain heavy elements, as the absolute PV energy is predicted to scale as $\sim Z^5$.\cite{Zel1977,Bast2011} These criteria can guide the initial selection of the molecular candidates. However, to evaluate the size of PV shifts in various vibrational transitions and to select the optimal transitions for measurements, explicit electronic structure calculations are necessary.

\begin{figure}
\begin{center}
\includegraphics[scale=0.06]{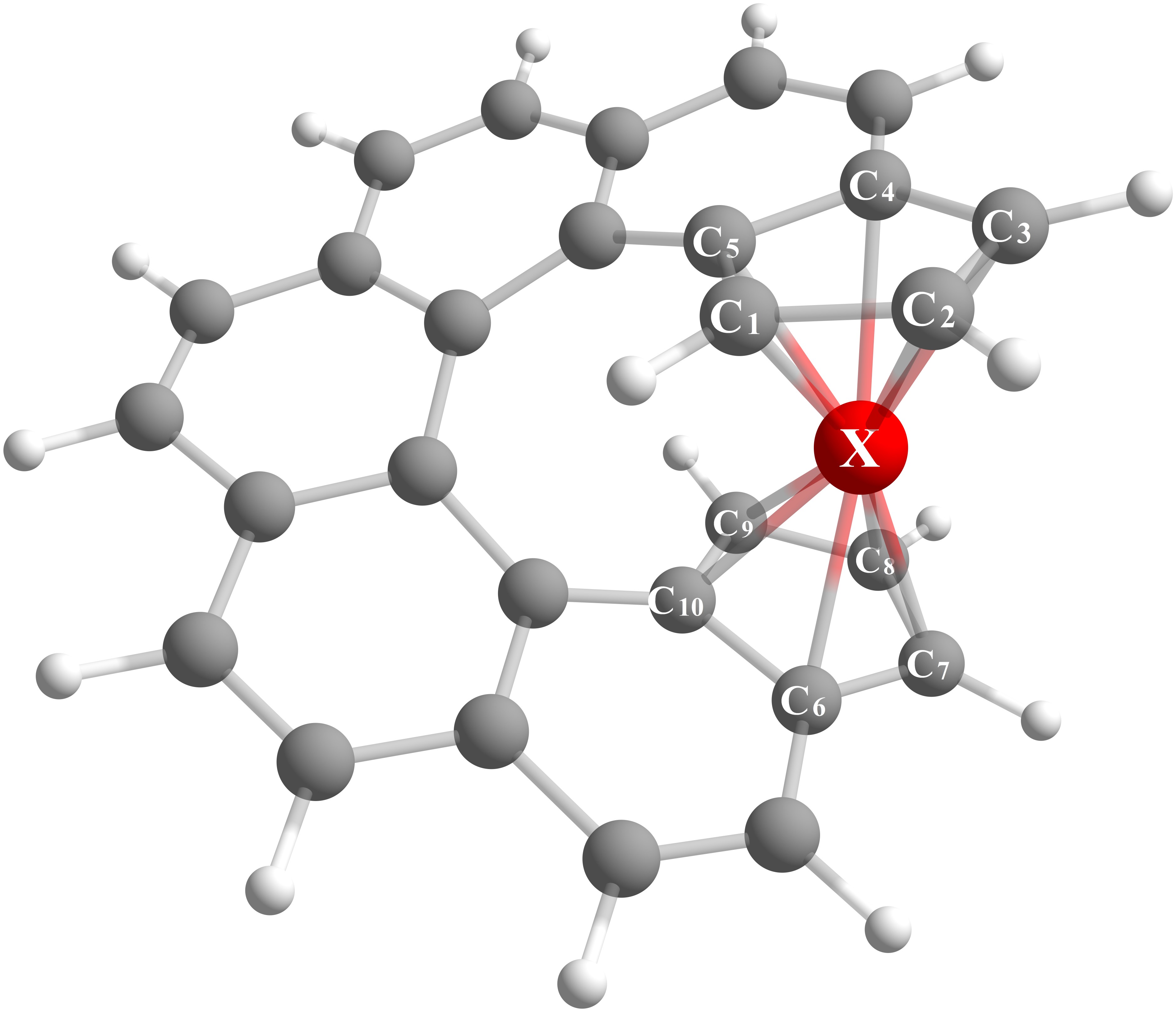}
\caption{Helical metalocene structure (X = Fe/Ru/Os) with 10 connected carbon atoms ($\eta$-10 complex).}
\label{fig:structure}
\end{center}
\end{figure}

In this work, we investigate helicenic organometallic metallocene-type complexes displayed in Figure \ref{fig:structure}, formed by metal coordination of a ligand having seven \emph{ortho}-fused aromatic rings including terminal cyclopentadienyl rings. These compounds exhibit inherent helical chirality. While the enantiopure forms of the heptahelicenic ligand may undergo partial racemization, especially at high temperatures,\cite{MarMar72,MARTIN1974347} complexation by a metal will increase the overall configurational stability.\cite{Johansson2009} Upon binding to Fe(II), helical ferrocene, an $\eta$-10 type complex, was readily synthesized by Katz and Pesti in 1982,\cite{Katz1982} characterized by X-ray crystallography in 1983,\cite{Dewan1983} prepared in enantioenriched form, and its chiroptical properties studied in 1986.\cite{sudhakar_asymmetric_1986} In principle, the preparation of heavier homologues, namely ruthenium and osmium complexes as depicted in Figure \ref{fig:structure}, may be considered. The size of the PV energy differences is expected to be amplified in these heavier compounds.\cite{Zel1977,Bast2011,FieHaaSal22}

We present a computational investigation of the parity-violating contributions to the vibrational transitions and nuclear magnetic resonance (NMR) shieldings of helical ferrocene, ruthenocene, and osmocene. We aim to determine whether the expected effects are large enough to be detectable in a modern, state-of-the-art experiment. We identify the vibrational transitions that are most promising for measurements, due to enhanced PV effects and practical considerations, such as high enough transition intensities and wavelengths that are easily accessible to current laser technologies. Progress in the development of the ultra-precise experiment dedicated to PV measurements is also presented.

\section{Theory and computational details}

The geometry optimization and the harmonic frequency analysis were carried out using the Q-Chem 5.2.2 program,\cite{qchem} within the $\omega$B97M-V density functional \cite{Mardirossian2016} combined with the Def2-TZVPP basis set,\cite{def2_basis} which includes triple-$\zeta$ valence (TZV) functions and large polarization functions (PP), together with the Stuttgart--Dresden small-core scalar-relativistic ECP$n$MWB pseudopotentials replacing the inner $n=28,\ 60$ electrons for Ru and Os, respectively.\cite{Def2_Pseudopotential}

\subsection{PV in vibrational transitions}

We investigated the PV effects for vibrational modes 9--129 for helical osmocene (Table~S1 in ESI$^\dag$), and compared selected modes 102, 103, and 108 with helical ruthenocene and ferrocene. For each mode, energy calculations were performed along the normal mode at 11 evenly spaced points $q_i$ between --0.5 \AA~ and 0.5 \AA~ (relative to the equilibrium structure) to obtain the potential energy curve $V(q)$, using the same level of theory as the geometry optimization.

We have also calculated the PV energy contributions at the same geometries along the normal mode, using the nuclear spin-independent (NSI) PV Hamiltonian as derived from the Z boson exchange between the electrons and nucleons:\cite{hpv_deriv,Bast2011}
\begin{equation} \label{eq:HamiltonianPV}
\hat{H}^\text{NSI-PV} = \frac{G_\mathrm{F}}{2\sqrt{2}} \sum_A%^\text{nuclei}
Q_{\mathrm{w},A} \sum_i%^\text{electrons}
\gamma_i^5 \rho_A(\vec{r}_i).%\vec{R}_A-\vec{r}_i).
\end{equation}
Here, $G_\mathrm{F}=2.22255\times 10^{-14}$ a.u. is the Fermi coupling constant, $Q_{\mathrm{w},A}=(1-4\sin^2\theta_\mathrm{W})Z_A-N_A$ is the weak charge of nucleus $A$, $\theta_\mathrm{W}$ is the Weinberg mixing angle (with $\sin^2\theta_\mathrm{W}=0.2319$ value adopted in DIRAC19),\cite{Montanet1994,DIRAC19} and $Z_A$ and $N_A$ are the proton and neutron numbers of nucleus $A$, respectively. The sums over indices $A$ and $i$ run over all nuclei and electrons, respectively. We also define $\gamma^5=i\gamma^0\gamma^1\gamma^2\gamma^3$, and $\rho_A(\vec{r})$ denotes the nuclear charge density of nucleus $A$ normalized such that $\int \rho_A(\vec{r}) d^3r = 1$. % to unity.

% When $\hat{H}^\text{NSI-PV}$ is treated as a perturbation within first-order perturbation theory, the PV energy shift at a given point along the normal coordinate $q$ is given by the expectation value  $E^\text{PV}=\bra{\Psi^\text{el}}H^\text{NSI-PV}\ket{\Psi^\text{el}}$,
% provided the wave function $\ket{\Psi^\text{el}}$ is either an eigenfunction of the unperturbed Hamiltonian within Born-Oppenheimer approximation or a stationary solution under orbital rotations, which holds for Hartree-Fock or density functional theory (DFT) solutions

Treating $\hat{H}^\text{NSI-PV}$ as a first-order perturbation, the PV energy shift---at a given point along the normal coordinate $q$---is given by the expectation value $E^\text{PV}=\bra{\Psi^\text{el}}H^\text{NSI-PV}\ket{\Psi^\text{el}}$, which holds for stationary Hartree--Fock (HF) or density functional theory (DFT) solutions.\cite{Hellmann37,Feynman39} The $E^\text{PV}$ values are obtained using the X2C/AMFI Hamiltonian\cite{Hess1996,AMFI}. In particular, the CAM-B3LYP* DFT functional\cite{ThiRauSch10}, which is optimized for PV calculations, combined with the augmented uncontracted dyall.v3z\cite{dyallv3z} basis set for the heavy metals and dyall.v2z\cite{dyallv2z} for the other atoms were used. These calculations were performed using the DIRAC19 program\cite{DIRAC19,dirac-paper}.

\begin{figure}
\begin{center}
\includegraphics[scale=0.43]{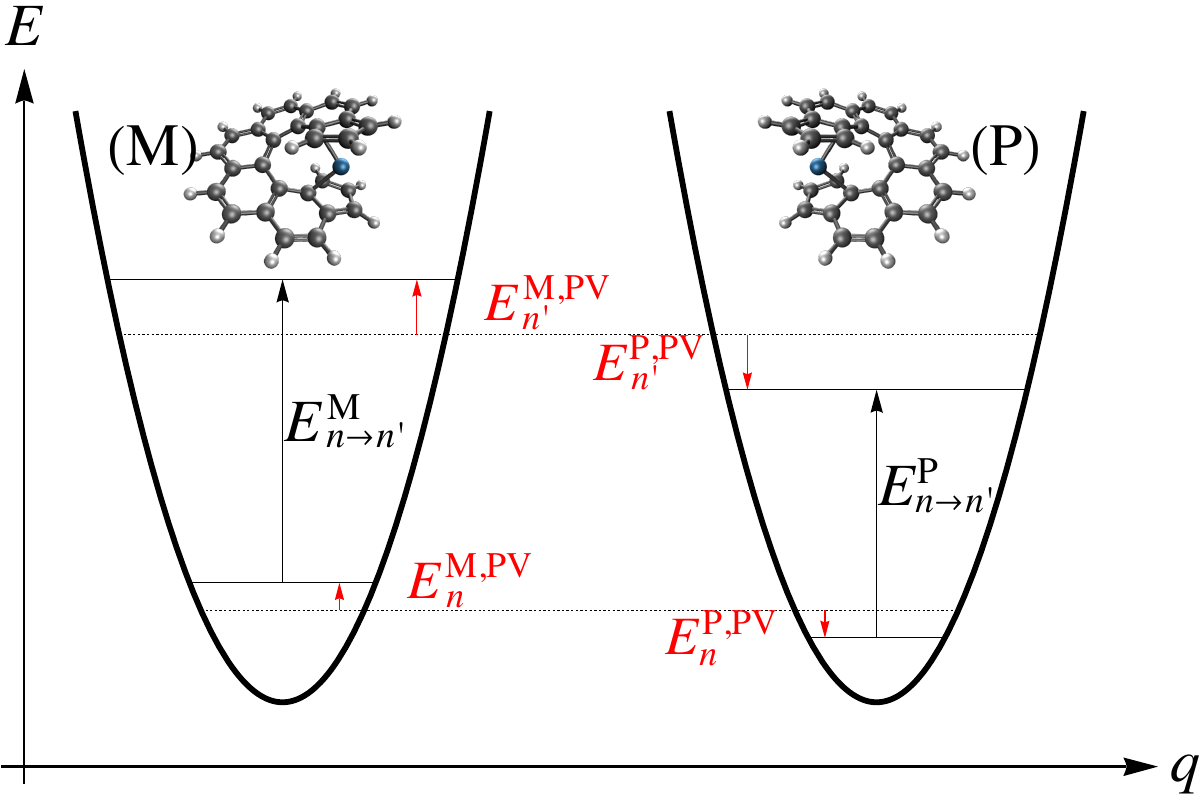}
\caption{Minus (M) and plus (P) enantiomers of helical osmocene and corresponding PV contributions to the vibrational transitions.}
\label{fig:RS_well}
\end{center}
\end{figure}

To obtain the PV shift for a specific vibrational mode, we used the numerical Numerov--Cooley (NC) procedure.\cite{numerov1,Cooley_1961,Bast_2017} Both the potential energy $V(q)$ and the PV energy contribution $E^\text{M,PV}(q)$ for the minus enantiomer (M) were fitted to a polynomial up to the sixth order. For a given normal mode, the vibrational wavefunction of the $n$-th vibrational energy level, $\ket{n}$, was obtained numerically and this was subsequently used for the calculation of PV energy contribution to the vibrational energy level ($E^\text{M,PV}_n$):
\begin{equation} \label{eq:E_pv_int}
  E^\text{M,PV}_n=\bra{n}E^\text{M,PV}(q)\ket{n}.
\end{equation}
The PV energy shift for a vibrational transition $n\rightarrow n'$ is then given by 
$E_{n\rightarrow n'}^\text{M,PV} = E_{n'}^\text{M,PV}-E_{n}^\text{M,PV}$.
Since the PV energy shifts  for a given vibrational level $n$ of the M and P enantiomers are related through $E_{n}^\text{M,PV}=-E_{n}^\text{P,PV}$ (Figure~\ref{fig:RS_well}), the difference between the respective transition energies is
\begin{equation} \label{eq:vib_diff}
\Delta E^\text{PV}_{n\rightarrow n'}=E_{n\rightarrow n'}^\text{M,PV}-E_{n\rightarrow n'}^\text{P,PV}=2E_{n\rightarrow n'}^\text{M,PV}.
\end{equation}
This energy difference is presented in terms of PV frequency shift by $\Delta\nu^\text{PV}_{n\rightarrow n'}=\Delta E^\text{PV}_{n\rightarrow n'}/h$, where $h$ is the Planck constant. The frequency shift divided by transition frequency ($\nu$) from harmonic frequency analysis ($\Delta \nu^\text{PV}_{n\rightarrow n'}/ \nu$) is an important theoretical predictor for the viability of measurement of PV in the corresponding transition.

\subsection{PV in NMR shielding constants}

We also analyzed the effect of parity violation on the NMR shieldings in the three molecules.
% in order to explore whether the sensitivity of the metalo-helicenes to these effects arising from weak interactions is also important for molecular properties other than energies.
%
The nuclear spin-dependent (NSD) PV contributions to the four-component (4C) NMR shielding tensors can be expressed in terms of linear response functions, which, for the M and P enantiomers of a chiral molecule, can be written as\cite{Laubender2003,Bast2006,Aucar-PV2023}
\begin{equation}
 \overleftrightarrow{\sigma}_A^\textrm{M/P,PV} = \frac{G_F\,m_p}{2\sqrt{2}\,\hbar\,c} \; \frac{\kappa_A}{g_A} 
 \langle\!\langle \; \rho_A(\vec{r}) \; c \vec{\alpha} \; ; \;  \vec{r}_\textrm{GO} \times c \vec{\alpha} \; \rangle\!\rangle^\textrm{M/P}, \label{eq:sigma-PV}
 \end{equation}
where we have defined $\kappa_A = -2\lambda_A \left(1-4\, \textnormal{sin}^2\theta_W\right)$, with $\lambda_A$ being a nuclear state-dependent parameter that we set to unity in our calculations to facilitate comparison with previous works.\cite{Laubender2003,Bast2006,Aucar-PVSR2021,Aucar-PVSR2022}
In addition, $m_p$ is the proton mass, $\hbar=h/(2\pi)$ is the reduced Planck constant, $c$ is the speed of light in vacuum, $\vec{\alpha} = \left( \alpha^1, \alpha^2, \alpha^3 \right)$ is the vector of $4 \times 4$ Dirac matrices (with $\alpha^i=\gamma^0 \gamma^i$ for $i=1,2,3$), $g_A$ is the nuclear $g$-factor of nucleus $A$, and $\vec{r}_\textrm{GO} = \vec{r} - \vec{R}_\textrm{GO}$ is the electron's position operator relative to a fixed gauge origin (GO) for the external magnetic potential.

When the two enantiomers (M and P) of a chiral molecule are analyzed, a difference will appear between the isotropic absolute shielding constants $\sigma^\textrm{M/P}_{A,\text{iso}}=\frac{1}{3}\textnormal{Tr}(\overleftrightarrow{\sigma}^\textrm{M/P}_A)$ of the same nucleus $A$ in the two enantiomers. In analogy to Eq.~\eqref{eq:vib_diff}, this difference will be given by\cite{Aucar-PVSR2022}
\begin{eqnarray}
|\Delta \sigma_{A,\text{iso}}|=|\Delta \sigma^\text{PV}_{A,\text{iso}}|=|\sigma^\text{M,PV}_{A,\text{iso}} - \sigma^\text{P,PV}_{A,\text{iso}}|
= 2 \, |\sigma^\text{M,PV}_{A,\text{iso}}|.
\end{eqnarray}
%
%since $\sigma^\text{S/R}_{A,\text{iso}}$ can be expressed as the sum of two terms: a parity-conserving contribution (arising from electromagnetic and weak interactions), which is identical for both enantiomers, and a parity-violating contribution (originating from weak interactions), which has the same magnitude but opposite sign for the two enantiomers.
Therefore, the measurable absolute value of the NMR frequency splitting of nucleus $A$, $\Delta \nu^\text{PV-NMR}_A$, in a static and homogeneous magnetic field of flux density $B_0$ is given by
\begin{eqnarray}\label{eq:freq-split}
%|\Delta \nu_A| = |\nu_A^{S} - \nu_A^{R}| = B_0 \, \frac{e}{4\pi \, m_p} |g_A \, \Delta \sigma_{A,\text{iso}}| ,
|\Delta \nu^\text{PV-NMR}_A| = B_0 \, \frac{e}{4\pi \, m_p} |g_A \, \Delta \sigma_{A,\text{iso}}| ,
\end{eqnarray}
where $e$ is the elementary charge.

The values of the diagonal tensor elements of $\overleftrightarrow{\sigma}_A^\text{PV}$ were computed using the DIRAC23 program package\cite{DIRAC23,dirac-paper}, employing the Dirac--Coulomb Hamiltonian. Given the computational demands of calculating the relativistic 4C linear response functions of Eq.~\eqref{eq:sigma-PV}, especially in large systems such as those studied in this work, the $\sigma_{A,\text{iso}}^\text{M/P,PV}$ constants can only be calculated using small basis sets. Although this restriction does not allow a highly accurate study of the property, these calculations do provide a reliable estimate of the expected order of magnitude of the effect.\cite{Aucar-PVSR2021,Aucar-PVSR2022} In particular, the dyall.aae2z basis sets were used for Fe, Ru, and Os,\cite{dyall_gomes_unpublished,dyall_4d_2007,dyallv3z} while the 3-21G basis sets were used for C and H.\cite{321g} Furthermore, the common gauge origin (CGO) prescription was applied, with the gauge origin of the external magnetic potential located at the molecular center of mass. The linear response functions were calculated at the random phase approximation (RPA) level of theory, employing DFT wavefunctions. In this way, electron correlation effects were taken into account. In particular, the hybrid functional PBE0 \cite{PBE0} was used, since it provides results in good agreement with experimental nuclear spin-rotation constants, a property closely related to NMR shieldings.\cite{Aucar-PVSR2021,Aucar-PVSR2022,Agus_NChDE_RSC2018,Aucar_CH3X}

%The nuclear charge density distribution was modeled using spherically symmetric Gaussian-type functions\cite{Visscher1997}. The explicit calculation of the (SS$\mid$SS) integrals was avoided by following the standard procedure on DIRAC\cite{Visscher1997-SSSS}. In all the cases, the basis sets for the small components were generated from the large component basis sets by applying the unrestricted kinetic balance prescription\cite{GAA_JCP99}. Besides, the response equations on Eq.~\eqref{eq:sigma-PV} were solved with respect to the property gradient associated with the operator $\vec{r}_\textrm{GO} \times c \vec{\alpha} $.

\section{Experimental considerations}

\subsection{Chemical synthesis}

The preparation of the heavier ruthenium and osmium homologues of the helical ferrocene (see Figure~\ref{fig:structure}) should be practically feasible by: (i) reproducing the synthesis of the helical ligand as described in the literature;\cite{Katz1982,Dewan1983} (ii) proceeding with enantiomeric resolution using classical chiral high-performance liquid chromatography (HPLC) techniques; (iii) incorporating heavy metal centers, Ru(II) or Os(II), through efficient synthetic methodologies.\cite{fischer_uber_1959,bobyens_crystal_1986,albers_chemistry_1986}
Besides synthetic feasibility, the chemical stability of such compounds is unknown and will need to be examined in detail, especially upon heating to bring it into the gas phase for the ultrahigh-resolution spectroscopic studies. Furthermore, the potential high toxicity of the possibly volatile osmium derivatives requires proper precautions.

\subsection{Ultra-precise mid-infrared spectroscopy of cold chiral molecules}

For measuring PV vibrational frequency differences between enantiomers, we envision at LPL a state-of-the-art, precise spectroscopy instrument based on the latest advances in cold-molecule research and frequency metrology. We will make a slow, cold beam of the organometallic molecules by producing them inside a buffer gas cell cooled to $\sim1$~K. This will provide the low temperature, low speed, and high intensity needed for measuring PV. We have already developed the methods needed to bring into the gas-phase and buffer-gas cool other organometallic species to a few kelvins.\cite{Cournol_2019,FieHaaSal22,Tokunaga2017,Asselin2017,darquie_valence-shell_2021} The molecular beam will be probed in a Ramsey interferometer\cite{shelkovnikov_stability_2008} based on frequency-stabilized metrology-grade lasers calibrated to the cesium fountain clocks of Laboratoire Temps Espace (LTE, the French time-frequency national metrology institute), which realize the standard of frequency of the international system of units. This machine will be sensitive enough to measure tiny changes in the vibrational frequency, and we project a $10^{-15}$ relative sensitivity ($<100$~mHz) on the frequency difference,\cite{Cournol_2019} a two-orders-of-magnitude improvement over the CHFClBr measurement.

When selecting vibrational modes for measurements, a number of experimental considerations have to be taken into account: both the magnitude of the PV frequency shifts and the intensity of the selected mode should be high enough to make the measurement feasible, and their frequencies should be accessible to metrology-grade laser technologies. At LPL, we have developed a method to stabilize mid-infrared lasers to an ultra-stable near-infrared optical frequency comb (OFC) referenced to primary frequency standards of LTE \emph{via} the REFIMEVE optical-fibre infrastructure~\cite{cantin_accurate_2021,refimeve}. Key aspects of the setup that require, in particular, non-linear mixing of the mid-infrared and the comb radiation in a crystal of AgGaSe$_2$ are briefly presented in the ESI.$^\dag$ This approach offers unparalleled sub-Hz frequency accuracies, stabilities, and laser line widths necessary for measuring the tiniest PV frequency differences. We have demonstrated this technology in the 1000~cm$^{-1}$ spectral window first using CO$_2$ lasers\cite{chanteau_mid-infrared_2013} and then quantum cascade lasers (QCLs)\cite{argence_quantum_2015,santagata_high-precision_2019,tran_near-_2024,tran_extending_2025}. The CO$_2$ lasers, until recently one of the very few available ultra-stable sources for precise mid-infrared spectroscopy, span only small portions of the 810--1120~cm$^{-1}$ range. Mid-infrared QCLs, on the other hand, are commercially available in the 800--2000~cm$^{-1}$ range and available more sporadically down to $\sim550$~cm$^{-1}$.

We have thus recently started to extend the spectral coverage of our metrology-grade QCLs/OFC system, specifically targeting the 550--600~cm$^{-1}$ and 1500--1600~cm$^{-1}$ regions, which, together with the 1000~cm$^{-1}$ window, host the most promising normal modes of the helical osmocene (see below). We have evaluated the spectral performance of the world's first $\sim580$~cm$^{-1}$ continuous-wave distributed feedback (DFB) QCL and demonstrated the first absorption spectroscopy at this wavelength using a QCL.\cite{manceau_demonstration_nodate,wang_wavelength_2025} We have even more recently acquired a 1560~cm$^{-1}$ QCL and carried out preliminary investigations of its frequency noise and emission line width, which we report in ESI.$^\dag$ Our current setup's spectral coverage is limited to the 850-1200~cm$^{-1}$ by the combination of the non-linear AgGaSe$_2$ crystal used and OFC emission spectrum. Frequency stabilizing QCLs at 550-600~cm$^{-1}$ and 1500-1600~cm$^{-1}$ requires to determine new such combinations.
% Reported demonstrations in the 1600~cm$^{-1}$ region have employed orientation-patterned (OP) GaAs [ref] and OP-GaP [ref] crystals as the nonlinear medium. However, at lower frequencies, studies are exceedingly rare given the dearth of sources.
We have thus carried out simulations of the non-linear mixing efficiency between the OFC and QCLs at our target wavelengths in a panel of birefringent and orientation-patterned (OP) crystals, testing both sum- or difference-frequency-generation (SFG or DFG) as the non-linear phenomenon, in order to identify options which maximize non-linear mixing efficiency. Details of this theoretical study are given in the ESI.$^\dag$ Our investigations indicate SFG in GaSe crystals as the solution of choice at both 580~cm$^{-1}$ and 1560~cm$^{-1}$. If cutting the GaSe crystal to the correct phase-matching angle is not possible, SFG in OP-GaAs, OP-GaP and AgGaSe$_2$ also happen to be reasonable fallback solutions.

The QCLs that will be used in the PV measurements typically cover a few wavenumbers only. As will be shown in the following, the size of the PV vibrational shift can vary by a few orders of magnitude depending on the vibrational mode. Theoretical guidance on the optimal vibrational modes for measurements is thus crucial for determining a target frequency range and designing the metrology-grade laser system to be built and tuned at LPL.

\section{Results and discussion}

\subsection{Optimized geometries}

The structure of helical ferrocene has been elucidated by the X-ray diffraction.\cite{Dewan1983} Our optimized structure is in good agreement with experiment, with root mean square deviation (RMSD) of 0.13 {\AA} (see also Figure~S1 in ESI$^\dag$ for the visual comparison). We expect similar accuracy for the optimized geometries of the two heavier systems, where experimental information is not yet available.

\subsection{PV in vibrational transitions}

\begin{table}[]
\caption{Calculated wavenumbers, intensities, PV frequency shifts, and total metal-carbon displacement for selected normal modes in helical osmocene promising for PV measurements. Modes of interest in the currently available sub-Hz metrology-grade laser window are given in bold font. *Mode 108 is listed for comparison in  Figures~\ref{fig:shift_comparison}--\ref{fig:6s5p} and the corresponding discussion below. }
\label{tab:recommendation}
\resizebox{\columnwidth}{!}{%
\begin{ruledtabular}
\begin{tabular}{rrrrrr}

Mode & $\tilde{\nu}$& Intensity & $\Delta \nu^\text{PV}_{0\rightarrow 1}$  & $\Delta \nu^\text{PV}_{0\rightarrow 1}/\nu$ &$d_\text{MC}$\\
number & (cm$^\text{--1}$)& (km/mol) & (Hz)  &&(\AA)\\
\hline
30& 586.5& 28.89& 1.03&5.84~x~10$^\text{--14}$&0.80
\\
31& 595.9& 3.55& --5.77&--3.23~x~10$^\text{--13}$&1.25
\\
32& 602.3& 1.18& 6.85&3.79~x~10$^\text{--13}$&1.46
\\
35& 636.0& 0.81& 2.64&1.38~x~10$^\text{--13}$&0.75
\\
36& 641.5& 7.82& --3.27&--1.70~x~10$^\text{--13}$&1.44
\\
 37& 658.3& 0.71&--3.04&--1.54~x~10$^\text{--13}$&0.87
\\
 40& 707.7& 0.86&2.46&1.16~x~10$^\text{--13}$&1.08
\\
 46& 795.0& 4.88&2.12&8.89~x~10$^\text{--14}$&0.54
\\
 48& 824.0& 16.58&--0.60&--2.44~x~10$^\text{--14}$&0.34
\\
 50& 836.9& 12.64&0.19&7.46~x~10$^\text{--15}$&0.24
\\
 54& 859.6& 18.37&--1.40&--5.44~x~10$^\text{--14}$&0.77
\\
 \textbf{56}& \textbf{867.0}& \textbf{73.99}&\textbf{0.37}&\textbf{1.41~x~10$^\text{--14}$}&\textbf{0.37}
\\
 \textbf{59}& \textbf{891.6}& \textbf{20.07}&\textbf{0.49}&\textbf{1.84~x~10$^\text{--14}$}&\textbf{0.50}
\\
\textbf{ 61}& \textbf{913.1}& \textbf{4.39}&\textbf{--2.83}&\textbf{--1.03~x~10$^\text{--13}$}&\textbf{0.49}
\\
 \textbf{72}& \textbf{1053.7}&\textbf{ 12.32}&\textbf{ --0.63}&\textbf{ --2.00~x~10$^\text{--14}$}&\textbf{0.62}
\\
 94& 1391.1& 9.87& --0.32& --7.66~x~10$^\text{--15}$&0.94
\\
 102& 1481.9& 5.55& 3.82& 8.60~x~10$^\text{--14}$&1.33
\\
 103& 1487.7& 8.34&--0.53&--1.19~x~10$^\text{--14}$&1.35
\\
 *108& 1592.5& 5.54& --0.09& 1.82~x~10$^\text{--15}$&0.29\\ 

\end{tabular}%
\end{ruledtabular}
}
\end{table}

\begin{figure}[]
\begin{center}
  \includegraphics[scale=1]{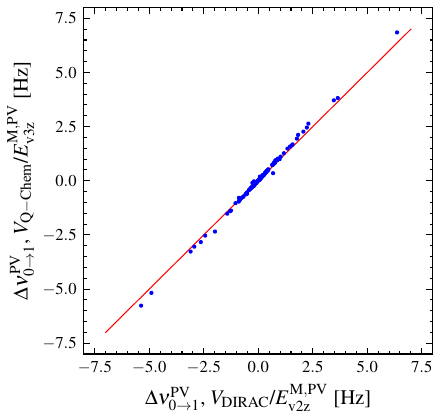}
  \caption{Correlation between PV frequency shifts in helical osmocene calculated on two different levels of theory -- the higher level uses dyall.v3z basis set for Os in the $E^\text{M,PV}$ calculations and potential $V(q)$ calculated in Q-Chem; the lower level uses dyall.v2z basis set for Os in the $E^\text{M,PV}$ calculations and potential $V(q)$ from the same DIRAC calculations. In both cases, the v2z basis set was used for the carbons and the hydrogens. Red line shows the ideal linear correlation.}
  \label{fig:basis_comparison}
\end{center}
\end{figure}

Table \ref{tab:recommendation} presents a selection of recommended vibrational modes of helical osmocene for potential PV measurements (Table~S1 in ESI$^\dag$ contains all calculated vibrational modes).
These modes were selected based on their frequency falling within the desired laser frequency range (500--2000~$\text{cm}^{-1}$) and on having exceptionally high intensity and/or high PV frequency shift. All these modes have a relative PV frequency shift $\Delta\nu_{\mathrm{PV}}/\nu$ of $\sim10^{-14}$ or more. This is a level of frequency sensitivity already demonstrated at LPL\cite{DarStoZri10,shelkovnikov_stability_2008} and at least an order of magnitude larger than the $10^{-15}$ sensitivity aimed for in the new experiment. It can be compared to the predicted $-8 \times 10^{-17}$ relative PV shift of CHFClBr\cite{quack3,peter1,thierfelder,rauhut,bruck} on which much of the experimental work has been conducted so far. Normal modes 56, 59, 61 and 72 are in the laser window with currently available sub-Hz metrology-grade sources at LPL, are reasonably infrared active, and have a predicted relative PV shift on the $10^{-14}$ level or more. Vibrational modes 56 and 61 look particularly promising, the former given its remarkably high intensity (the highest intensity transition in the entire spectrum), the latter given its very large predicted $\sim10^{-13}$ relative PV shift. Importantly, this is twice higher than the most sensitive transition of Os(acac)$_3$ reported in our previous study\cite{FieHaaSal22} in the 1000~cm$^{-1}$ region where our sub-Hz laser technology is currently available. Table \ref{tab:recommendation} also lists a series of vibrational modes exhibiting relative PV shifts ranging from $\sim10^{-14}$ to a few $10^{-13}$ and reasonable intensities in the 600~cm$^{-1}$ and 1500~cm$^{-1}$ spectral windows in which metrology-grade laser systems are currently being developed at LPL. Vibrational modes 30 and 32 are especially favorable, not only because they exhibit respectively a particularly intense transition and the highest predicted relative PV shift in the entire spectrum -- about $4\times10^{-13}$ -- but also because of their lower 600~cm$^{-1}$ frequencies, which may prevent the onset of intramolecular vibrational energy redistribution that could obscure the spectra at higher frequencies.

\begin{figure}[]
\begin{center}
  \includegraphics[scale=0.56]{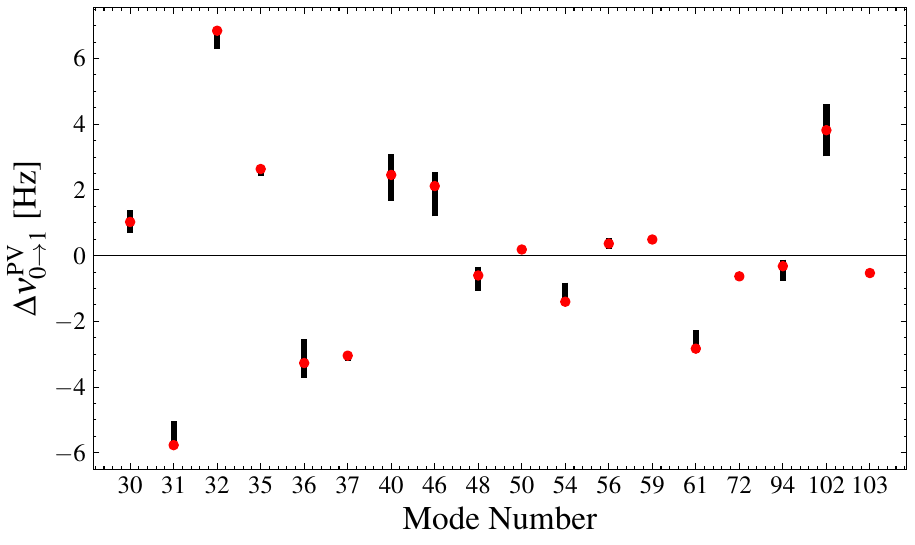}
  \caption{PV Frequency shifts of the recommended modes calculated using different DFT functionals and Hartree--Fock. Red dots represent the CAM-B3LYP* values, while black bars represent the range of frequency shifts calculated using all methods (based on the values in Table S3 in ESI$^\dag$).}
  \label{fig:dft_comparison}
\end{center}
\end{figure}

In order to assess the robustness of the predicted relative PV shifts with respect to the computational parameters, we have carried out additional calculations. 
In Figure~\ref{fig:basis_comparison}, we compare PV shifts of all calculated modes at the production level of theory using the dyall.v3z basis set\cite{dyallv3z} for Os to a lower-level calculation using a more modest dyall.v2z basis set\cite{dyallv2z} as well as potential energy curves from the DIRAC calculation instead of the original Q-chem potentials, consistent with the harmonic frequency analysis. We found the obtained vibrational frequency shifts in excellent agreement with the higher-level calculations, indicating that this lower-cost approach (at least twice as fast) can be used as a predictor in future work.
Furthermore, we have investigated the effect of different DFT functionals on the calculated PV frequency shifts and found the results to be in good agreement for the CAMB3LYP*\cite{ThiRauSch10}, CAMB3LYP*(EFG)\cite{CamB3LYPEFG}, B3LYP\cite{B3,LYP,LDA1,B3LYP}, PBE\cite{PBE}, LDA\cite{LDA1,LDA2} functionals, and for Hartree--Fock calculations, as shown in Figure~\ref{fig:dft_comparison}, confirming the robustness of our predictions. The complete list of the calculated PV contributions for these modes can be found in Table~S3 in ESI.$^\dag$ 
Note that functionals CAMB3LYP, CAMB3LYP* and CAMB3LYP*(EFG) differ in their attenuation parameters which are optimized, respectively, for energy calculations ($\alpha$=0.19, $\beta$=0.46, and $\mu$=0.33), PV calculations ($\alpha$=0.2, $\beta$=0.12, and $\mu$=0.9), and electric field gradient (EFG) calculations ($\alpha$=0.4, $\beta$=0.179, and $\mu$=0.99).

\begin{figure}[]
\begin{center}
  \includegraphics[scale=1]{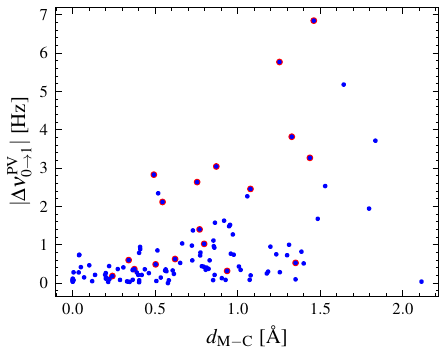}
  \caption{Absolute values of the PV frequency shifts of the vibrational transitions of helical osmocene plotted against the total metal-carbon displacement. The vibrations selected as promising for measurements (see Table~\ref{tab:recommendation}) are marked in red.}
  \label{fig:os_all}
\end{center}
\end{figure}

Our previous study on Ru(acac)$_3$\cite{FieHaaSal22} identified a positive correlation between the size of the PV frequency shift for a specific vibration and the total ligand displacement along the vibration motion. Here, we perform a similar analysis, defining the  metal-carbon displacement as
\begin{equation} \label{eq:ligand_displacement}
 d_\text{MC} = \sum_{i=1}^{10} \sqrt{(\Delta x_{\text{MC},i})^2 + (\Delta y_{\text{MC},i})^2 + (\Delta z_{\text{MC},i})^2},
\end{equation}
where the summation runs over the 10 neighboring cyclopentadienyl carbon atoms (see Figure~\ref{fig:structure}) and $\Delta \alpha_{\text{MC},i}\ (\alpha=x,y,z)$ represent the relative displacements of the involved metal and carbon atoms.
The PV frequency shifts and the ligand displacements for all calculated vibrational modes of helical osmocene are given in Table~S1 in ESI$^\dag$. The calculated absolute PV shifts are plotted against the total ligand displacement of the corresponding transition in Figure \ref{fig:os_all}. We observe that small ligand displacements with respect to the metal consistently yield relatively low PV frequency shifts. However, conversely,  large ligand displacements do not necessarily lead to a high PV frequency shift, making the displacement a predictor with limited applicability for this class of molecules. 

To aid our discussion further, we consider three representative vibrational modes with frequencies within the range of commercially available lasers currently considered at LPL and with a relatively high intensity. 
These are mode 102 (cyclopentadienyl symmetric breathing), mode 103 (cyclopentadienyl asymmetric breathing), and mode 108 (side ring deformation). Modes 102 and 103 form a symmetric pair, in which mode 102 exhibits a significantly higher PV shift compared to mode 103. While modes 102 and 103 both have very large ligand displacements (1.33 \AA~and 1.35 \AA, respectively), mode 108 with its relatively small ligand displacement (0.29 \AA) is selected for contrast.

\begin{figure}[]
\begin{center}
  \includegraphics{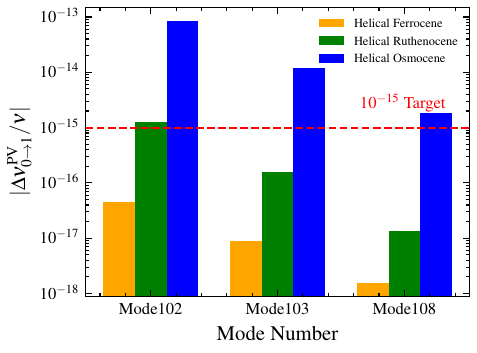}
  \caption{Relative PV frequency shifts of vibrational modes 102, 103, and 108 for helical ferrocene, helical ruthenocene, and helical osmocene.}
  \label{fig:shift_comparison}
\end{center}
\end{figure}

Figure~\ref{fig:shift_comparison} shows the magnitude of relative PV frequency shifts for the three selected vibrational modes in helical ferrocene, ruthenocene, and osmocene. 
The structure and properties of the three systems are expected to be similar, and thus the same modes were selected in each molecule. 
Of the three systems, only the heaviest, helical osmocene, has relative PV frequency shifts that surpass the projected detection limit of  $10^{-15}$ on all three modes, motivating us to focus further on this molecule.
As expected from the small ligand displacement, mode 108 yields a small PV frequency shift (--0.09 Hz). In contrast, mode 102, with the highest PV frequency shift (3.83 Hz) in the 1000~cm$^{-1}$ spectral window currently accessible at LPL, has a seven times higher PV shift than mode 103 (--0.53 Hz), despite a similar magnitude of ligand displacement. It is thus clear that other factors besides the ligand displacement play a role in determining the size of the PV frequency shift. 

\begin{figure}[h]
\begin{center}
  \includegraphics[scale=1.0]{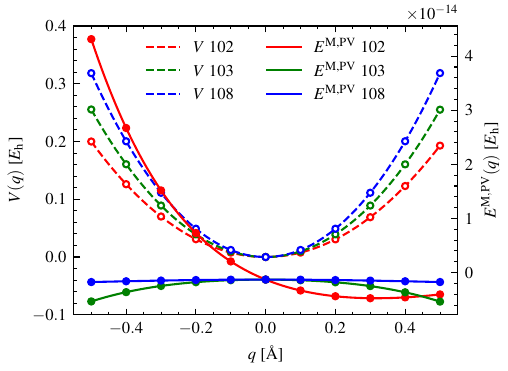}
  \caption{PV energy curves (solid lines, filled circles) and potential energy curves (dashed lines, empty circles) for vibrational modes 102, 103, and 108 in helical osmocene.}
  \label{fig:pv_v_plot}
\end{center}
\end{figure}

Figure \ref{fig:pv_v_plot} shows the potential energy curves and the PV energy curves for the three vibrational modes discussed above. While the potential energy curves are almost ideally harmonic and very similar for all three modes, the shapes of the parity-violating energy curves are markedly different. Furthermore, based on the analytic perturbative contribution analysis in ESI$^\dag$ Table~S1, PV shifts of these three modes are dominated by the curvature term of PV curve with the anharmonic term being negligible. In agreement with the ligand displacement analysis, the PV curve of mode 108 is almost flat and close to zero.
However, while the magnitude of the ligand displacement for modes 102 and 103 (which form an antisymmetric pair) is almost identical, mode 102 has a significantly larger PV response compared to mode 103, leading to a significantly higher PV frequency shift.
This observation is in line with the relatively poor correlation between PV shift and atom displacement in Figure~\ref{fig:os_all}.

\begin{figure}[ht!]
\begin{center}
\includegraphics[width=0.45\textwidth]{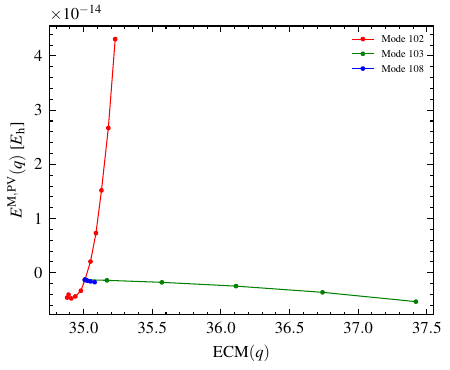}
\caption{$E^\text{M,PV}$ vs ECM for vibrational modes 102, 103, and 108 in helical osmocene. Modes 103 and 108 are symmetric, and thus the positively and negatively displaced points are overlaid, resulting in seemingly fewer data points.}
\label{fig:SI_ECM102}
\end{center}
\end{figure}

We have analyzed vibrational modes 102, 103, and 108 using the electronic chirality measure (ECM)\cite{2003.Bellarosa} obtained using the PyECM software.\cite{pyECM24} ECM is a descriptor based on the electronic density that quantifies the chirality degree of a molecule.\cite{2003.Bellarosa} It was shown that for some organic molecules, such as alanine and glyceraldehyde, there is a correlation between $E^\text{M,PV}$ and ECM.\cite{2024.Aucar_ECM} 
Figure \ref{fig:SI_ECM102} shows the relationship between the total PV energy and the value of ECM along the displacement coordinate for the three selected vibrational modes. 
In general, we observe a correlation between the  ECM and the $E^\text{M,PV}$ values within each mode. However, the behavior for each of the studied modes is vastly different. For mode 108, the minimal change in the ECM is consistent with the very small change in $E^\text{M,PV}$ along the displacement. On the other hand, the relationships between the ECM and $E^\text{M,PV}$ values in modes 102 and 103 are markedly different. While mode 102 displays a big change in $E^\text{M,PV}$, the corresponding ECM values remain within a very tight range. In contrast, the large ECM variation along mode 103 accompanies a rather small $E^\text{M,PV}$ change.
For the three modes, the higher magnitude for $E^\text{M,PV}$ is obtained within the range of higher ECM.
Furthermore, ECM values are by definition unsigned, and hence the changes of sign commonly observed for $E^\text{M,PV}$ values (as seen, e.g., in mode 102) cannot be directly reproduced.
From this behavior, it can be concluded that the ECM analysis in this system does not allow for the prediction of which mode will present the higher relative PV frequency shift.

\begin{figure}[h]
\begin{center}
  \includegraphics[scale=1.0]{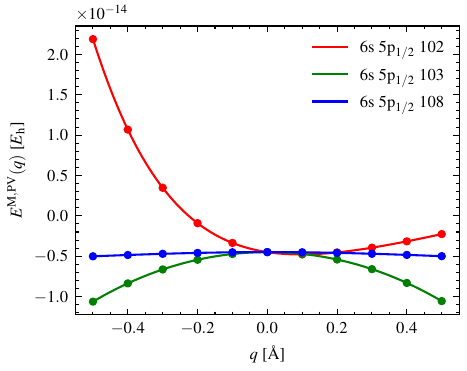}
  \caption{Osmium intraatomic $\langle 6\text{s}|5\text{p}_{1/2}\rangle$ contributions to PV energy curves for vibrational modes 102, 103, and 108 in helical osmocene.}
  \label{fig:6s5p}
\end{center}
\end{figure}

In a further attempt to interpret the results, we employed chirality density analysis\cite{Bast2011,Hegstrom1991} and orbital projection analysis,\cite{Faegri2001,Dubillard2006} details of which can be found in ESI$^\dag$ Section~S5. Chirality density plots (Figures~S3 and S4 in ESI$^\dag$) reveal that significant changes in chiral density along the vibrational coordinate $q$ occur at the Os nucleus in mode 102, leading to a strong PV response, while in modes 103 and 108 these changes are smaller and located mostly outside of the Os nucleus and penetrating it only partially, leading to relatively flat $E^\text{M,PV}$ curves.
The atomic orbital projection analysis\cite{Faegri2001,Dubillard2006} shows that the PV curve qualitatively follows the largest osmium $\langle 6\text{s}|5\text{p}_{1/2}\rangle$ contribution (compare Figures~\ref{fig:pv_v_plot} and~\ref{fig:6s5p}).
In ESI$^\dag$ Section~S5.3, we discuss the details of the projection analysis and complications due to the significant influence of molecular polarization effects.

%The motivation behind the search for various predictors of the size of the PV frequency shift is that such predictors could potentially remove the need for the computationally demanding explicit relativistic calculations on the candidate molecules and allow selection of promising candidates based on either already known molecular properties (such as the displacements of the atoms around the heavy metal) or much cheaper calculations, such as in the case of ECM. Alternatively, one can test the possibility of performing significantly less expensive explicit calculations of the PV frequency shifts, as we also did above. 

\subsection{PV effects in NMR shielding constants}

To further characterize the behavior of PV effects in the three molecules investigated in this work, we also evaluated their contributions to NMR shielding constants. At a magnetic flux density of $B_0 =$~20~T, as considered in recent experimental proposals for searches of PV effects on NMR parameters,\cite{Eills2017} the predicted NMR frequency splittings range from about 0.02 mHz for Fe in ferrocene to about 0.8 mHz for Os in osmocene (see Table~\ref{tab:PV-shield}). For helical osmocene, this value lies within roughly one order of magnitude of the projected experimental sensitivity. Although still challenging, this suggests that the system may be of interest for PV searches also in the NMR context and could motivate further studies using other spectroscopic approaches.

%At a magnetic flux density of $B_0 =$~20~T, such as the one used in recent experimental proposals for searches of PV effects on NMR parameters\cite{Eills2017}, the NMR frequency splitting reaches values ranging from about 0.02~mHz for Fe in ferrocene to about 0.8~mHz for osmium in osmocene, as seen in Table~\ref{tab:PV-shield}. For osmocene, this result is just one order of magnitude below the projected sensitivity of these types of experimental searches\cite{Eills2017}.

\begin{center}
\begin{ruledtabular}
\begin{table}[h!]
\caption{Calculated values of $2\,|g_A\,\sigma^\text{M,PV}_{A,\text{iso}}|$ in parts per trillion (ppt) and $|\Delta \nu^\text{PV-NMR}_A|$ (for $B_0=20~\text{T}$, in mHz) for the $A$ nuclei (with $A=$ Fe, Ru, and Os) in the helicene series of molecules.}
\label{tab:PV-shield}
\centering

\begin{tabular*}{\linewidth}{@{\extracolsep{\fill}} lcc}
$A$ & $2\,|g_A\,\sigma^\text{M,PV}_{A,\text{iso}}|$ / ppt & $|\Delta \nu^\text{PV-NMR}_A (20~\text{T})|$ / mHz \\
\hline 
  $^{57}$Fe & 0.11 & 0.02 \\
 $^{101}$Ru & 0.59 & 0.09 \\
 $^{189}$Os & 5.38 & 0.82
\end{tabular*}
\end{table}
\end{ruledtabular}
\end{center}

\section{Conclusions}

We have calculated the PV vibrational frequency shifts for a selection of normal modes in helical osmocene, a highly promising candidate species for the first detection of parity violation in molecules. This molecule is the first system with inherent helical chirality investigated in the context of PV searches. We have pinpointed a number of mid-infrared transitions which fall in accessible laser windows and exhibit exceptionally large PV shifts of hundreds of millihertz to $\sim7$~Hz. This pushes the $|\Delta\nu_{\mathrm{PV}}/\nu|$ of these modes into the $\sim10^{-13}$ regime, exceeding by up to 2 orders of magnitude the projected instrumental sensitivity of the ultrahigh-resolution experiment being built at LPL. %For transitions for which proper laser technologies are readily available, it is a tenfold improvement over our previous study on Os(acac)$_3$\cite{FieHaaSal22}. 

The effect of computational setting on the calculated PV energy shifts was investigated, and a lower-cost theoretical approach was identified for future studies. The results were also found to be robust with respect to the used DFT functional. A variety of proposed structural and electronic predictors of PV effects were tested. In this class of systems, we found their applicability to be limited, motivating the use of explicit calculations based on low-cost computational methods.

We have also investigated the size of the PV effects in NMR shielding constants. For helical osmocene, these were found to be just an order of magnitude below the projected sensitivity of planned experimental searches for PV in NMR spectroscopy. 

The predicted large PV effects in helical osmocene provide a strong motivation to synthesize this compound and to bring it into the gas phase. A strategy for synthesizing helical osmocene has been proposed. Once synthesized, its enantiomeric resolution and stability upon heating to bring it into the gas phase for the ultrahigh-resolution spectroscopic studies will be investigated.

\section*{Acknowledgments}
The authors thank the Center for Information Technology of the University of Groningen for their support and for providing access to the H\'abr\'ok high-performance computing cluster. 
Eduardus wishes to acknowledge the Indonesia Endowment Fund for Education/\textit{Lembaga Pengelola Dana Pendidikan} \text{(LPDP)} for research funding. 
LFP acknowledges the support from the project number VI.C.212.016 of the talent programme VICI, financed by the Dutch Research Council (NWO), and support from the Scientific Grant Agency of the Slovak Republic (project 1/0254/24).
IAA acknowledges partial support from FONCYT through grants PICT-2021-I-A-0933 and PICT-2020-SerieA-0052, and CONICET through grant PIBAA-2022-0125CO. JJA acknowledges the Institute for Modelling and Innovative Technologies (IMIT, CONICET) and the National University of the Northeast (UNNE) for support and access to the IMIT high-performance computing cluster. The work of IAA and AB was supported by the project \textit{Probing Particle Physics with Polyatomic molecules} with project number OCENW.M.21.098 of the research programme M2, which is financed by the Dutch Research Council (NWO). The work of ABo, MM and BD was supported by the Agence Nationale de la Recherche projects Ultimos (Grant No. ANR-24-CE30-2959-01); CNRS; Universit\'{e} Sorbonne Paris Nord; and was part of 23FUN04 COMOMET that has received funding from the European Partnership on Metrology, co-financed by the European Union's Horizon Europe Research
and Innovation Programme and from by the Participating States, funder ID: 10.13039/100019599. ABo, MM, and BD thank Anne Amy-Klein for the first SFG calculations based on AgGaSe$_2$, preliminary modeling of OP non-linear crystals, and Anne Amy-Klein, Christof Janssen, and Laurent Hilico for many fruitful discussions. In memory of Jean-Jacques Zondy.

% \nocite{*}
\bibliography{citation}% Produces the bibliography via BibTeX.

\end{document}